\documentclass{iau} 

\usepackage{graphicx}

\newcommand{\Vc}        {v_{\rm c}}
\newcommand{\fDM}        {f_{\rm DM}}
\newcommand{\fP}        {f_{\rm P}}

\newcommand{\df} {DF}
\newcommand{\HI} {{H{\sc{i}}}}
\newcommand{\vJ} {\mathrm{\textbf{J}}}

\newcommand{\e} {\mathrm{e}}

\def\Gyr{\,{\rm Gyr}}

\title[Self-consistent Modelling of the Milky Way using Gaia data] 
{Self-consistent Modelling of the Milky Way using Gaia data}

\author[David R. Cole \& James Binney]   
{David R. Cole$^1$ \and James Binney$^2$}

\affiliation{$^1$Rudolf Peierls Centre for Theoretical Physics, Keble Road, Oxford, \\ OX1 3NP, United Kingdom \\ email: {\tt david.cole@physics.ox.ac.uk} \\[\affilskip]
$^2$Rudolf Peierls Centre for Theoretical Physics, Keble Road, Oxford, \\ OX1 3NP, United Kingdom \\email: {\tt binney@physics.ox.ac.uk}}

\pubyear{2017}
\volume{330}  
\jname{Astrometry and Astrophysics in the Gaia sky}
\editors{eds. Recio-Blanco, de Laverny \& Brown.}
\begin{document}

\maketitle

\begin{abstract}
Angle/action based distribution function (\df ) models can be optimised
based on how well they reproduce observations thus revealing the
current matter distribution in the Milky Way. Gaia data combined with
data from other surveys, e.g. the RAVE/TGAS sample, and its full
selection function will greatly improve their accuracy.

\end{abstract}

\firstsection 
\section{Introduction}

Our knowledge of how baryons were accreted by galaxies such as the
Milky Way is limited however but unprecedented amounts of data are
becoming available from large scale surveys. It is vital that we use
these data to improve our understanding of the local group and an
excellent starting point is to find the current distribution of matter
in the Galaxy. We cannot observe directly the distribution of dark
matter so we must use the observable components as tracers to build
dynamical models from which we can discover the distribution of dark
matter. If we assume the Galaxy is in statistical equilibrium, we
can exploit Jeans theorem (\cite[Jeans 1916]{Jeans1916}) and presume
that the distribution function (\df) $f(x,v)$ is a function of
integrals of motion $I(x,v)$ only. There is an infinite choice of
integrals because any function of the integrals is also an integral
but the best choices are the actions $J_i$, which can be uniquely
complemented by canonically conjugate variables to make a complete set
($\theta$,J) for phase-space coordinates.

\section{Modelling Process}

The Milky Way can be built up from its components; the dark halo, thin
and thick discs, stellar halo, bulge and gas disc. Each of these
components is modelled either by a distribution function,
$f(\textbf{J})$ (first three components above), or a fixed
potential. By making a sensible estimate for the parameters of these
components we can make an estimate for the total potential. We then
use the St\"ackel Fudge (\cite[Binney
  2012,2014]{Binney2012,Binney2014}) to find the actions and by
integrating over velocity:
\begin{equation}
\rho(x) = \int d^3v f(\textbf{J}(\textbf{x},\textbf{v}))
\end{equation}
we estimate the density. We then solve Poisson's equation to find the
new potential and iterate this procedure until after a few iterations
the total potential converges.

\subsection{Disc \df}

The \df\ of the discs in our models is superposition of the
``quasi-isothermal'' components introduced by \cite[Binney \&
  McMillan (2011)]{Binney2011}. It has the form
\begin{equation}
\label{eq:qi}
 f(J_r,J_z,J_\phi)=f_{\sigma_r}(J_r,J_\phi)f_{\sigma_z}(J_z,J_\phi),
\end{equation}
where $f_{\sigma_r}$ and $f_{\sigma_z}$ are
\begin{equation}
\label{planeDF}
 f_{\sigma_r}(J_r,J_\phi)\equiv
 \frac{\Omega\Sigma}{\pi\sigma_r^2\kappa}
      [1+\tanh(J_\phi/L_0)]\e^{-\kappa J_r/\sigma_r^2},
      f_{\sigma_z}(J_z,J_\phi)\equiv\frac{\nu}{2\pi\sigma_z^2}\,\e^{-\nu
        J_z/\sigma_z^2}.
\end{equation}
The thick disc is represented by a single quasi-isothermal \df, while
the thin disc's \df\ is built up with a quasi-isothermal for each
coeval cohort of stars. The velocity-dispersion parameters depend on
$J_\phi$ and the age of the cohort. The star-formation rate in the
thin disc decreases exponentially with time, with characteristic time
scale $t_0=8\Gyr$. In addition a parameter $F_{\rm thk}$ controls the
fraction of mass contributed by the thick disc.

\subsection{Dark halo \df}

Our dark halo \df\ is based on the form introduced by \cite[Posti et
  al. (2015)]{Posti2015} which in isolation self-consistently
generates a density distribution which has a NFW (\cite[Navarro, Frenk
  \& White 1997]{Navarro1997}) profile (equation
\ref{eq:Posti}). This \df\ is a function of $h(\textbf{J})$ which is
an almost linear function (a homogeneous function of degree unity) of
the actions $J_i$. The haloes generated are isotropic centrally and
mildly radial when $r>r_s$. The anisotropy can be changed by varying
the linear function of the $J_i$. These haloes closely resemble the
haloes formed in dark-matter-only simulations. Specifically
 \begin{equation}
\label{eq:Posti}
\fP(\vJ)=\frac{N}{J_0^3}\frac{(1+J_0/h)^{5/3}}{(1+h/J_0)^{2.9}}
\end{equation}
The scale action $J_0$ encodes the scale radius around which the slope
of the radial density profile shifts from $-1$ at small radii to $-3$
far out. From equation (\ref{eq:Posti}) it follows that $f_P(\vJ)\sim
|\vJ|^{-5/3}$ as $|\vJ|\rightarrow0$.

\subsection{Observational constraints}

We use several sets of observations to constrain the parameters of
the \df s at various stages of the modelling. The observations include
the astrometry of H$_2$O and SiO masing stars (\cite[Reid et
  al. 2014]{Reid2014}), the distribution of radio-frequency lines of
\HI\ and CO emission in the longitude-velocity plane, the stellar parameters and
distance estimates in the fourth RAVE data release (\cite[Kordopatis
  et al. 2013]{Kordopatis2013}) (see Section \ref{sec:kine}) and
finally the vertical density profile from SDSS. We assume that the
population from which the RAVE sample is drawn is identical to that
studied by \cite[Juri\'{c} et al. (2008)]{Juric2008}.

\section{Results}

\begin{figure}[b]
\begin{center}
  \includegraphics[width=3.6in]{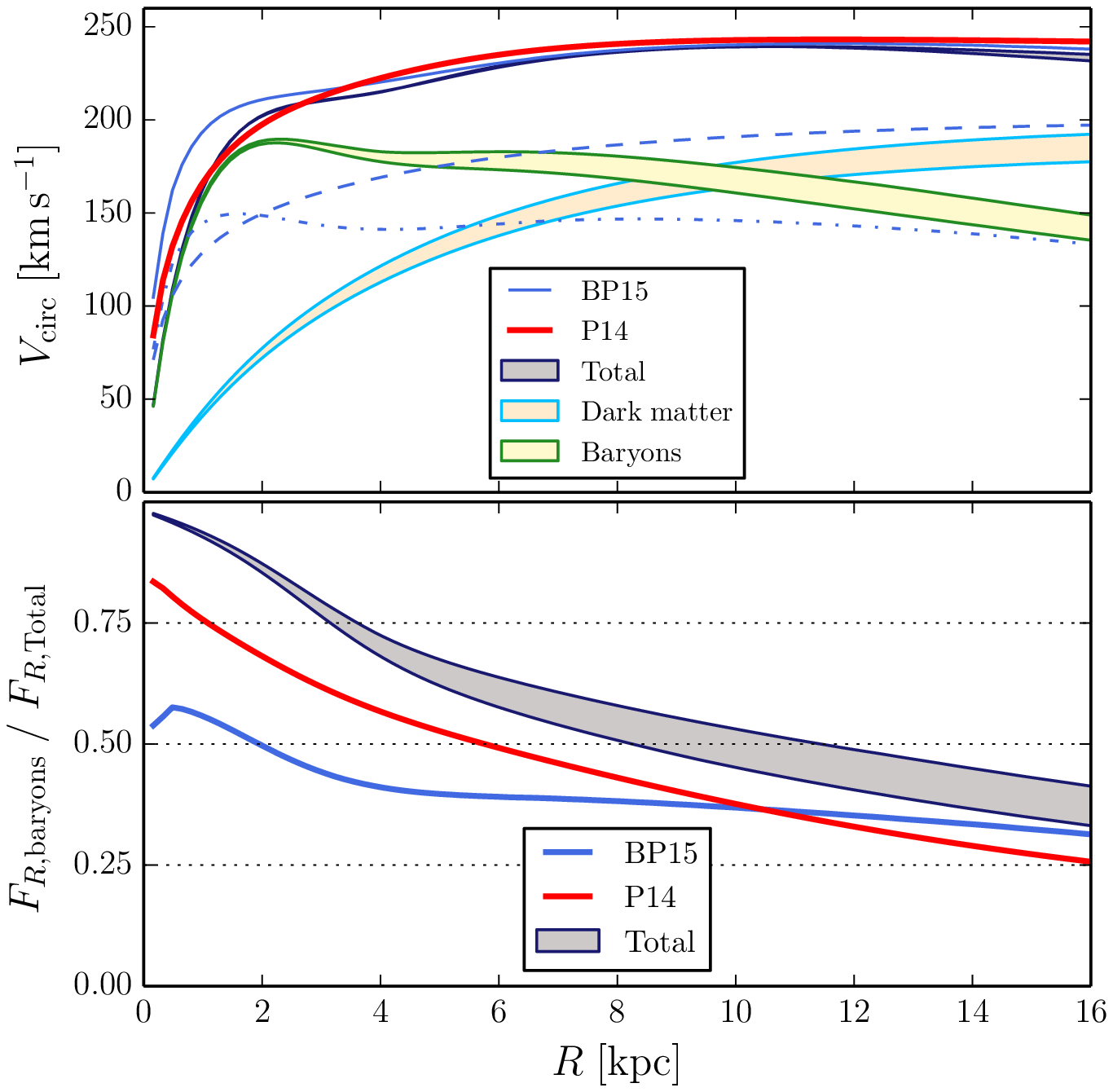} 
 \caption{Upper panel: $\Vc(R)$ for one of the Galaxy models of
   \cite{Cole2017} (dark grey line) compared to the models of
   \cite{Binney2015} (BP15 black line) and \cite[Piffl et
     al. 2014]{Piffl2014} (P14 light grey line). The range of dark
   halo and baryonic contributions for \cite{Cole2017} are shown as
   filled areas (baryonic in foreground). The dashed line is the dark
   halo contribution to the rotation curve of BP15 and the dash-dotted
   line is its baryonic component.  Lower panel: the ratio of radial
   forces from the baryonic component to the total mass distribution
   for the same models shown above with \cite{Cole2017} shown as a
   filled area. }
   \label{fig1}
\end{center}
\end{figure}

\cite[Piffl et al. (2014)]{Piffl2014} used RAVE data and SDSS Juric
(2008) data to constrain the mass of DM within solar radius, R$_0$.
The dark halo was included as a potential not a \df. \cite[Binney \&
  Piffl (2015)]{Binney2015} used the \df\ in equation \ref{eq:Posti}
for the dark halo in their self-consistent model of the Milky Way. In
this model the halo was in self-consistent equilibrium with the other
components so it had been adiabatically compressed by the baryons from
its original NFW form. Hence it mimicked a scenario where baryons
accumulated quiescently in the Galaxy's dark halo. An NFW halo becomes
more centrally concentrated, and with so much dark matter at low radii
(Fig.  \ref{fig1}) the matching disc contains too few stars to satisfy
the microlensing data (\cite[Popowski et
  al. 2005]{Popowski2005}). This implies that the infinite phase-space
density of particles at $J=0$ characteristic of an NFW \df\ does not
survive the accretion of baryons. The baryons cannot have accumulated
entirely adiabatically but the most tightly bound dark matter
particles were upscattered.  This scattering of DM particles by
baryons reduces their phase-space density and had greatest impact near
$J=0$.

In order to model this scattering process \cite[Cole \& Binney
  (2017)]{Cole2017} modified the NFW \df\ by setting
$f(\textbf{J})=g(h)f_{NFW}(h)$ with $g\sim{h}^{5/3}$ for small $h$ and
$g\sim1$ at large $h$. This shifts particles from very low $J$ to
higher $J$.  The functional form for $g$ is
\begin{equation}
\label{eq:defsg}
g(h)=\left[{h_0^2\over h^2}-\beta{h_0\over h}+1\right]^{-5/6},
\end{equation}
 where $h_0$ is an arbitrary constant with the dimensions of action
 that sets the scale of the almost constant-density core of the
 \df\ and $\beta$ is determined by requiring the total mass of dark
 matter is conserved. The free parameters in $\fDM$ are $J_0$, which
 sets the NFW scale radius, $h_0$, which sets the size of the dark
 halo's core, and the normalisation $N$.

\cite[Cole \& Binney (2017)]{Cole2017} found (see Fig. \ref{fig1})
that with this modified dark halo \df\ their models of the Milky Way
have a similarly good fit to the observations but the central regions
are now dominated by baryons. The dark matter fraction is reduced and
consistent with the results of surveys of microlensing events. In
summary they found the local DM density $\rho_{DM}{\gtrsim}0.012$
$M_\odot$ pc$^{-3}$, stellar disc scale radius $R_d\sim2.9$ kpc and
stellar disc mass M$_d\gtrsim4\times10^{10}$ $M_\odot$.

\section{Using RAVE/TGAS}
\label{sec:kine}

Our current modelling uses RAVE data in 8 spatial bins at $R_0\pm$1
kpc and at 0, 0.3, 0.6, 1.0, 1.5 kpc in $\left|z\right|$ to compute
velocity distributions predicted by the \df\ at the mean positions. In
order to take advantage of Gaia DR1 (\cite[Gaia Collaboration
  2016]{Gaia2016}) we are developing a method using the RAVE/TGAS
observations (\cite[Kunder et al. 2017]{Kunder2017}). In order to do
this we will develop a full selection function $S(s)$ for this
sample. Deriving an a priori calculation of $S(s)$ needs a full
chemodynamical model of the MW disc $S(s,τ,[Fe/H])$ such as used in
\cite[Sch\"{o}nrich \& Bergemann 2014]{Schoenrich2014} and in addition
we need to model the exact distribution of stars in age and
metallicity in the solar neighbourhood. \cite[Sch\"{o}nrich and Aumer
  2017]{Schönrich2017} did this using the RAVE selection function of
\cite[Wojno et al. (2017)]{Wojno2017} and found that at fixed
metallicity $S(s,\tau)$ falls off exponentially with scale 0.12 kpc at
$s>0.2$ kpc.

The selection function for TGAS is biased towards younger stars which
are more likely to be seen so the kinematics will appear cooler than
they really are. Our models need to take this into account. They
already have age but not metallicity and so we can add metallicity by
use of a suitable metallicity \df. Then we can compute the likelihoods
of our model based on the resulting selection function.

\section{Action based modelling software library}

Our modelling is currently being rewritten using AGAMA (Action-based
Galaxy Modelling Architecture) which is a library of low-level
programs containing interfaces and generic routines required to create
the functions described here. The main sets of functions include
gravitational potential and density interfaces, action/angle
interface, interface for creating gravitationally self-consistent
multicomponent galaxy models etc. The code can be downloaded from
https://github.com/GalacticDynamics-Oxford/Agama

\end{document}